\documentclass[journal,comsoc]{IEEEtran}

\usepackage[T1]{fontenc}
\makeatletter
\newcommand*{\rom}[1]{\expandafter\@slowromancap\romannumeral #1@}
\makeatother
\usepackage{cite}
\usepackage{bbding}
\usepackage{pifont}
\usepackage{amssymb}
\usepackage{amssymb}
\usepackage{enumerate}
\usepackage{textcomp}
\usepackage{graphicx}
\graphicspath{ {images/} }
\usepackage{makecell}
\usepackage{slashbox}
\usepackage{array}
\usepackage{booktabs}
\usepackage{tabularx}
\usepackage{subfig}
\usepackage{color, colortbl}
\definecolor{Gray}{gray}{0.9}
\usepackage{graphicx}
\usepackage{hyperref}
\usepackage{multirow}
\newcolumntype{L}{>{\centering\arraybackslash}m{3cm}}
\usepackage{tabularx}
\usepackage{url}
\usepackage{stfloats}
\usepackage{amsmath}
\usepackage{authblk}
\usepackage{multicol}
\usepackage[table]{xcolor}

\newcolumntype{g}{>{\columncolor{Gray}}c}
\ifCLASSINFOpdf
\else
\fi
\begin{document}

\title{A Survey on Sensor-based Threats to Internet-of-Things (IoT) Devices and Applications}
%
%
%


\author[1]{\small Amit Kumar Sikder}
\author[2]{\small Giuseppe Petracca}
\author[1]{\small Hidayet Aksu}
\author[2]{\small Trent Jaeger}
\author[1]{\small A. Selcuk Uluagac}

\affil[ ]{\footnotesize $^{1}$Cyber-Physical Systems Security Lab (CSL) \hspace{30mm} $^{2}$Systems and Internet Infrastructure Security Lab (SIIS)} 

\affil[ ]{\footnotesize \hspace{-6mm} Department of Electrical and Computer Engineering \hspace{24mm} Department of Computer Science and Engineering } 

\affil[ ]{\footnotesize Florida International University, Miami, Florida-33199, USA \hspace{16mm} Penn State University, University Park, Pennsylvania-16801, USA}

\affil[ ]{\footnotesize \hspace{-10mm} Email: \{asikd003, haksu, suluagac\}@fiu.edu \hspace{42mm} Email: \{gxp18, tjaeger\}@cse.psu.edu}

\maketitle
\thispagestyle{empty}
\pagestyle{plain}
\begin{abstract}
The concept of Internet of Things (IoT) has become more popular in the modern era of technology than ever before. From small household devices to large industrial machines, the vision of IoT has made it possible to connect the devices with the physical world around them. This increasing popularity has also made the IoT devices and applications in the center of attention among attackers. Already, several types of malicious activities exist that attempt to compromise the security and privacy of the IoT devices. One interesting emerging threat vector is the attacks that abuse the use of \textit{sensors} on IoT devices. IoT devices are vulnerable to \textit{sensor-based threats} due to the lack of proper security measurements available to control use of sensors by apps. By exploiting the sensors (e.g., accelerometer, gyroscope, microphone, light sensor, etc.) on an IoT device, attackers can extract information from the device, transfer malware to a device, or trigger a malicious activity to compromise the device. 
In this survey, we explore various threats targeting IoT devices and discuss how their sensors can be abused for malicious purposes. Specifically, we present a detailed survey about existing sensor-based threats to IoT devices and countermeasures that are developed specifically to secure the sensors of IoT devices. Furthermore, we discuss security and privacy issues of IoT devices in the context of sensor-based threats and conclude with future research directions. 
\end{abstract}

\begin{IEEEkeywords} Sensory side-channel attacks, CPS attacks, IoT threats, IoT device security.
\vspace{.3cm}
\end{IEEEkeywords}

%
\IEEEpeerreviewmaketitle

\section{Introduction}
\IEEEPARstart{I}{nternet of things} (IoT) is a concept that describes a network of interconnected devices which has advanced capabilities to interact with devices and also with human beings and its surrounding physical world to perform a variety of tasks \cite{bari2013internet}. In this context, the use of sensors on IoT devices ensures a seamless connection between the devices and the physical world. Indeed, modern IoT devices come with a wide range of sensors (e.g., accelerometer, gyroscope, microphone, light sensor, etc.) which enable more efficient and user-friendly applications \cite{5560598}. Using these sensors, IoT devices can sense any changes in their surrounding and take necessary actions to improve any ongoing task efficiently \cite{5606038}. The ability to sense changes in the physical world have made IoT devices able to make autonomous decisions, whereas, efficient communication between the devices and the physical world have made the IoT devices very popular in various application areas: from personal healthcare to home appliances, from big industrial applications to smart cities. IoT devices are in many possible application domain. The increasing popularity and utility of IoT devices in divergent application domains made the IoT industry to grow at a tremendous rate. According to a report by Business Insider, there will be 30 billion devices connected to the Internet by 2020 and more than 6 trillion dollars will be invested in manufacturing of IoT devices in the next five years \cite{IoTCPS}. 

The use of sensors in IoT devices inevitably increases the functionality of the devices; however, the sensors can also be used as vehicles to launch attacks on the devices or applications. For instance, recently, there have been several attempts to exploit the security of IoT devices via their sensors~\cite{203854, son2015rocking, nahapetian2016side}. Attackers can use the sensors to \textit{transfer malicious code} or \textit{trigger message} to activate a malware planted in an IoT device \cite{6654855, Hasan:2013:SCH:2484313.2484373}, \textit{capture sensitive personal information} shared between devices (e.g., smartphone, smartwatch, etc.) \cite{schlegel2011soundcomber, Zhuang:2009:KAE:1609956.1609959, Maiti:2015:WYT:2802083.2808397}, or even \textit{extract encrypted information} by capturing encryption and decryption keys \cite{del2015side}. These sensor-based threats can pose significant risk to the IoT systems and applications than the conventional attacks as the manufacturers are not fully aware of these threats yet \cite{6997498}. Indeed, sensor-based threats are becoming more prevalent with time because of the easy access to the sensors and limited security measures that consider these threats \cite{petracca2017aware, petracca2016agility, petracca2016aware, Petracca:2015:APA:2818000.2818005}. Furthermore, attackers do not need any complicated tools to access the sensors which makes sensor-based threats easier to execute \cite{schlegel2011soundcomber, placeraider}. Hence, trivial execution, easy access to the sensors, and lack of knowledge about the sensor-based threats constitute significant risks for the IoT devices and applications. Understanding these sensor-based threats is necessary for researchers to design reliable solutions to detect and prevent these threats efficiently.\par

\textbf{\textit{Contributions---}}In this paper, we provide a survey on threats that can be exploited to attack sensors in IoT devices and applications. The contributions of this paper are: 
\begin{itemize}
    \item \textit{First}, we present a detailed discussion about sensor management systems in various IoT operating systems and identify the important shortcomings of the existing systems. 
    \item \textit{Second}, we provide a detailed taxonomy of sensor-based threats in the IoT world. 
    \item \textit{Third}, we discuss existing security solutions for IoT devices and their shortcomings in the context of sensor-based threats.
    \item \textit{Fourth}, we articulate several open issues and discuss future research direction that could contribute to secure sensors in IoT devices and applications.
\end{itemize} \par
\textbf{\textit{Organization---}}The rest of the paper is organized as follows. We give the definition and general architecture of IoT devices in Section \rom{2}. In Section \rom{3}, we briefly discuss existing sensor management system of IoT OSes and their shortcomings in detecting sensor-based threats. We present a set of sensor-based threats in Section \rom{4}. In Section \rom{5}, we articulate security approaches that have been proposed to secure sensors of IoT devices. Future research in the area of sensor-based threats and security of IoT devices are described in Section \rom{6}. Finally, we conclude this paper in Section \rom{7}.
\section{Background: Components of IoT}
In this section, we identify the components of IoT devices as it is relevant to understand the significance of sensor-based threats on IoT devices and applications. In general, an IoT device can be explained as a network of things which consists of hardware, software, network connectivity, and sensors~\cite{Cyber1}. 
Hence, the architecture of IoT devices comprises four major components: sensing, network, data processing, and application layers (as depicted in Figure~\ref{IoT})~\cite{farooq2015critical}. A detailed description of these layers is given below.


\begin{table*}[htb!]
\centering
\arraybackslash
\fontsize{8}{10}\selectfont
\begin{tabular}{ccp{9cm}}

\toprule
\textbf{Sensor Type}    & \textbf{Sensor Name}   & \textbf{Description} \\ \hline
\midrule

\centering
                                     \multirow{1}{2cm}{\centering Motion Sensors}                & Accelerometer               & \textbullet \hspace{1mm}An electro-mechanical device which can measure changes in acceleration forces along x, y, and z-axis. \par
                                     \textbullet \hspace{1mm}Detects various types of motion like shake, tilt, etc. and adjusts the display of the device accordingly.    \\ \cline{2-3} 
                                                                              & Linear Acceleration Sensor & \textbullet \hspace{1mm}An accelerometer which can detect acceleration along one axis without considering the effect of gravitational force.\par \textbullet \hspace{1mm}Helps to adjust the display with motion.                                                                                        \\ \cline{2-3} 
                                                                                 & Gyroscope & \textbullet \hspace{1mm}Measures the rate of change of angular momentum in all three axes.\par
                                        \textbullet \hspace{1mm}Detects rotational movement of the device and adjusts display accordingly.                                         \\ \hline
\multirow{1}{2cm}{\centering Environmental Sensors} & Light Sensor                                           &  \textbullet \hspace{1mm}A photodiode which changes                             characteristics with the change of light intensity.                              \par \textbullet \hspace{1mm}adjusts brightness and contrast of the display of the device. \par \textbullet \hspace{1mm} Controls                                automatic lighting system.        \\ \cline{2-3} 
                                                                                 & Proximity Sensor & \textbullet \hspace{1mm}IR-based sensor to detect the presence of nearby objects without any physical contact. \par \textbullet \hspace{1mm}Reduces power consumption of the display by disabling the LCD backlight and avoids inadvertent touches. \\ \cline{2-3} 
                                                                                 & Temperature Sensor   & \textbullet \hspace{1mm}Measures temperature of the device as well as ambient temperature.\par \textbullet \hspace{1mm}Controls and sets the temperature in a device.                                                             \\ \cline{2-3} 
                                                                                 & Audio Sensor  & \textbullet \hspace{1mm}Two types of audio sensor: microphone and speaker. \par \textbullet \hspace{1mm}Microphone: Detects acoustic signal.\par \textbullet \hspace{1mm}Speaker: Playbacks audio signal.                                                                  \\ \cline{2-3} 
                                                                                 & Camera      & \textbullet \hspace{1mm}Deals with light intensity, device ambiance, etc. to capture pictures and videos of surroundings. \par \textbullet \hspace{1mm} Provides live video feeds.       \\ \cline{2-3}
                                                & Barometer  & \textbullet \hspace{1mm}Measures the pressure of the device peripheral. \\ \cline{2-3}
                                                & Heart rate  & \textbullet \hspace{1mm}Measures the heart rate of the user in beat per second. \\ \hline
\multirow{1}{2cm}{\centering Position Sensors}                                                 & GPS                       & \textbullet \hspace{1mm}Captures signal from the satellite to infer the location of the device. \par \textbullet \hspace{1mm}Helps in navigation systems.                                                  \\ \cline{2-3} 
                                                                                 & Magnetic Sensor     & \textbullet \hspace{1mm}Measures device's magnetic field with respect to earth's magnetic field. \par \textbullet \hspace{1mm}It is also used to fix display position by considering the magnetic field.               \\
\bottomrule                                                                    
\end{tabular}
\caption{Sensors available in most IoT devices.}
\label{sensors}
\end{table*}

\subsection{Sensing Layer}
The main purpose of the sensing layer is to identify any phenomena in the devices' peripheral and obtain data from the real world. This layer consists of several sensors. Using multiple sensors for applications is one of the primary features of IoT devices \cite{khan2012future}. Sensors in IoT devices are usually integrated through sensor hubs~\cite{perera2013dynamic}.  A sensor hub is a common connection point for multiple sensors that accumulate and forward sensor data to the processing unit of a device. A sensor hub uses several transport mechanisms (Inter-Integrated Circuit (I2C) or Serial Peripheral Interface (SPI)) for data flow between sensors and applications. These transport mechanisms depend on IoT devices and create a communication channel between the sensors and the applications to collect sensor data. Sensors in IoT devices can be classified in three broad categories as described below. A detailed description of various IoT sensors is given in Table~\ref{sensors}.

\subsubsection{Motion Sensors}
Motion sensors measure the change in motion as well as the orientation of the devices. There are two types of motions one can observe in a device: \textit{linear} and \textit{angular} motions. The linear motion refers to the linear displacement of an IoT device while the angular motion refers to the rotational displacement of the device. 

\subsubsection{Environmental Sensors}
Sensors such as Light sensor, Pressure sensor, etc. are embedded in IoT devices to sense the change in environmental parameters in the device's peripheral. The primary purpose of using environmental sensors in IoT devices is to help the devices to take autonomous decisions according to the changes of a device's peripheral. For instance, environment sensors are used in many applications to improve user experience (e.g., home automation systems, smart locks, smart lights, etc.). 

\begin{figure}[t!]
  \centering
    \includegraphics[height=6cm, width=6cm]{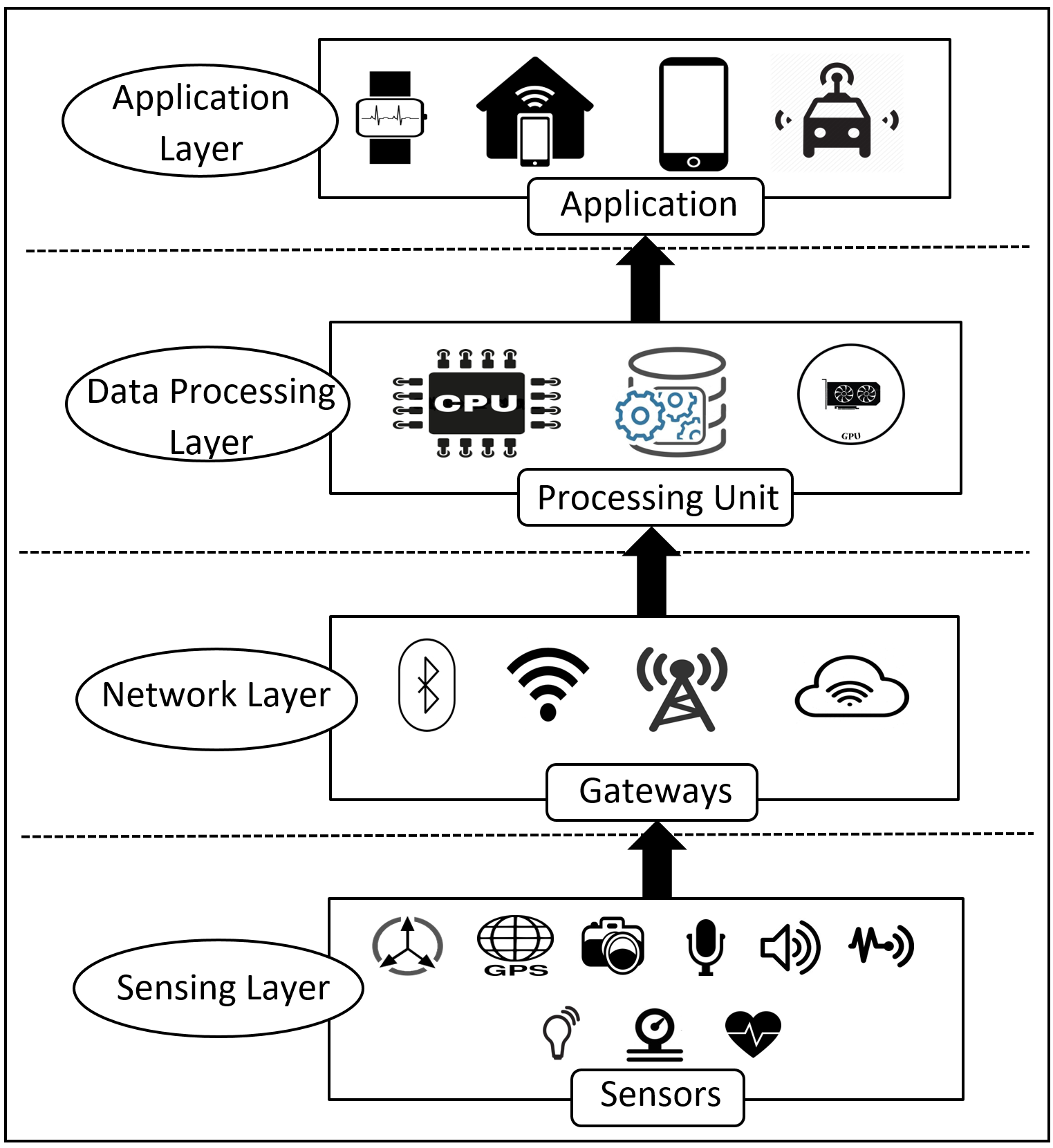}
      \caption{IoT Architecture Layers and Components.}
      \label{IoT}
\end{figure}

\subsubsection{Position sensors}
Position sensors of IoT devices deal with the physical position and location of the device. Most common position sensors used in IoT devices are magnetic sensors and Global Positioning System (GPS) sensors. Magnetic sensors are usually used as digital compass and helps to fix orientation of device display. On the other hand, GPS is used for navigation purposes in IoT devices. 

\subsection{Network Layer}
The network layer acts as a communication channel to transfer data, collected in the sensing layer, to other connected devices. In IoT devices, the network layer is implemented by using diverse communication technologies (e.g., Wi-Fi, Bluetooth, Zigbee, Z-Wave, LoRa, cellular network, etc.) to allow data flow between other devices within the same network.

\subsection{Data Processing Layer}
The data processing layer consists of the main data processing unit of IoT devices. The data processing layer takes data collected in the sensing layer and analyses the data to take decisions based on the result. In some IoT devices (e.g., smartwatch, smart home hub, etc.), the data processing layer also saves the result of the previous analysis to improve the user experience. This layer may share the result of data processing with other connected devices via the network layer.

\subsection{Application Layer}
The application layer implements and presents the results of the data processing layer to accomplish disparate applications of IoT devices. The application layer is a user-centric layer which executes various tasks for the users. There exist diverse IoT applications, which include smart transportation, smart home, personal care, healthcare, etc. \cite{Atzori20102787}.

\section{Sensor Management Systems in IoT OSes}
Emerging IoT systems create a many-to-many relationship between apps and sensors that OSes manage. Most IoT systems use more than one sensor to perform a task; thus, it is impractical to implement a standalone management system for each sensor. Moreover, to perform a task, an application usually needs to access data from multiple sensors. A separate sensor management for each sensor can cause delay in the data flow from the sensors to the application, which hampers user experience \cite{lane2010survey}. Hence, a sensor management system is needed to manage and ensure secure data acquisition from all the sensors. 
In this section, we discuss existing sensor management systems implemented by current IoT OSes. \par

\begin{figure}[t!]
  \centering

    \includegraphics[height=7cm, width=6.5cm]{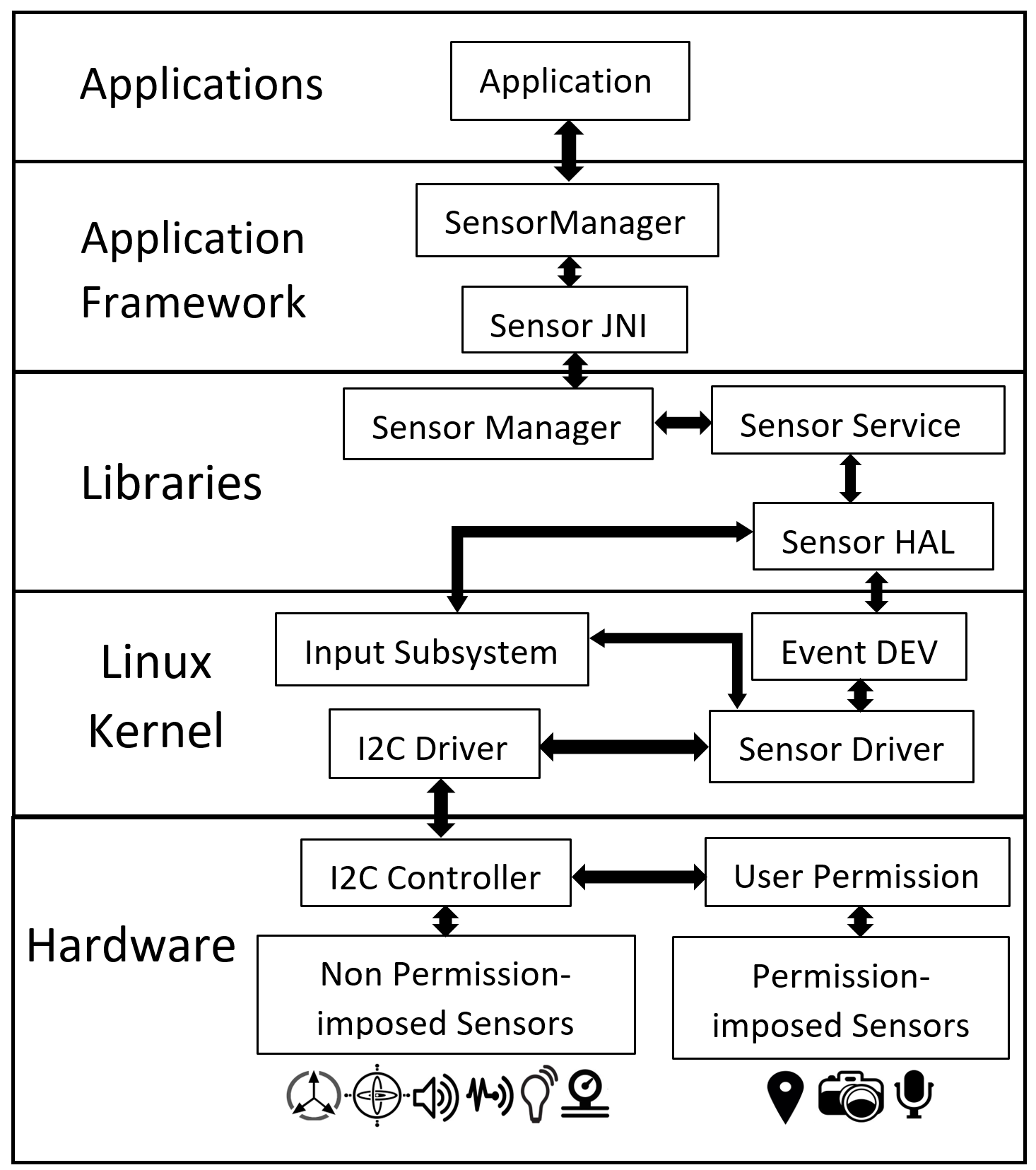}
      \caption{Example of Sensor Management System for Android.}
      \label{android}
\end{figure}

IoT devices may run one of a variety of OSes (i.e., Android, iOS, Windows Phone OS, Blackberry OS, etc.). Most of these OSes follow a permission-based sensor management system to control access and data flow between the applications and the sensors~\cite{iOS,Android}. As Android OS holds the highest market share in IoT domain (approximately 37\%), we briefly discuss Android sensor management system in this section~\cite{market}. A detailed overview of Android sensor management system is given in Figure~\ref{android}. Whenever an application wants to access a sensor in Android OS, it has to communicate via a sensor manager software. An application first sends a request to the sensor manager to register the desired sensor. This registration request includes the desired sensor parameters (e.g., frequency, delay, etc.) and the \textit{SensorEventListener} for the desired sensor. After receiving the request, the sensor manager creates a \textit{ListenerService} for the application and maps the request with the designed sensor driver to acquire sensor data. If more than one App requests access for the same sensor, sensor management system runs a multiplexing process to register one sensor to multiple Apps. This data acquisition path from the application to the sensor driver is initiated by the Hardware Abstraction Layer (HAL) in the Android OS as shown in Figure~\ref{android}. HAL mainly binds the sensor hardware with the device driver to acquire data. The sensor driver then activates the requested sensor and creates a data flow path from the sensor to the app~\cite{androidsen}. On the other hand, Windows and Blackberry OSes use Sensor Class Extension to connect sensor hardware with the device driver~\cite{winsen, blacksen}. Windows OS also uses User Mode Driver Framework to detect sensor access request and create a data acquisition path between sensor API and the APP. In iOS, the sensor management system is divided into four core services: Core Motion, Core Audio, Core Location, and Core Video \cite{iOSmotion}. The Core Motion service provides access to the motion sensors and some of the environmental sensors (e.g., barometer, light, proximity, etc.). The audio sensors (microphone and speakers), GPS, and the camera can be accessed via the Core Audio, the Core Location, and the Core Video services, respectively. These services provide data flow between sensors and apps according to the requests. 

Recently, Samsung introduced a new platform for smart devices named \textit{Samsung SmartThings} \cite{samsung}. This platform connects and controls all the IoT devices used in a home automation system. The sensor management systems of several devices can be controlled from one common platform (a hub or smartphone). Unlike other systems, Samsung SmartThings offers a capability-based sensor management system. With SmartThings, applications interact with smart devices based on their capabilities, so once the capabilities that are needed by a SmartApp are specified, and once the capabilities that are provided by an IoT device are identified, the devices - based on the device's declared capabilities - are selected for use within a specific SmartApp. 



The main shortcoming of the existing sensor management systems is the dependency on the user's consent for sensor access. Most OSes used in IoT devices rely on a permission-based access control for a specific subset of the supported sensors including camera, microphone, and GPS. Whenever an application is installed in an IoT device, it asks the users to grant permission to access various sensors (e.g., camera, microphone, and GPS). Thus, malicious applications may trick the user in granting access to sensitive sensors to launch sensor-based attacks \cite{schlegel2011soundcomber, placeraider, 203854}. Users are typically unaware of what the malicious applications are actually doing with the sensed data \cite{petracca2016aware, petracca2017aware}. Furthermore, permissions are imposed on selected sensors only (e.g., camera, microphone, and GPS); thus, applications can easily access other no-permission imposed sensors such as accelerometer, gyroscope, light sensor, etc., as discussed in the following sections in further details. These sensors can be exploited maliciously and various sensor-based threats (e.g., information leakage, denial-of-service, etc.) can be launched on IoT devices \cite{shen2015input, owusu2012accessory, Marquardt:2011:IDV:2046707.2046771}. Hence, existing sensor management systems of various IoT OSes are unable to prevent the abuse of sensors in IoT devices.



\section{Sensor-based Threats to IoT Devices}
As existing sensor management systems and security schemes cannot provide adequate security to the sensors, attackers can exploit these sensors in various ways. In this section, we provide a discussion about sensor-based threats on IoT devices and survey the existing attack scenarios confirmed by researchers \cite{schlegel2011soundcomber, placeraider, Petracca:2015:APA:2818000.2818005, petracca2016agility, petracca2016aware, petracca2017aware}. \par
In general, sensor-based threats refer to passive or active malicious actions which try to accomplish its malicious intents by exploiting the sensors. Sensor-based threats can be passive like observing the behavior of the device without obstructing the normal operations of the device or active such as false sensor data injection, transferring malicious sensor code to the device. Further, sensor-based threats in IoT devices can be categorized into four broad categories based on the purpose and nature of the threats: The categories are: \textit{(1) Information Leakage, (2) Transmitting Malicious Sensor Patterns or Commands, (3) False Sensor Data Injection,} and \textit{(4) Denial-of-Service}. These threats are described below.

\subsection{Information Leakage}

Information leakage is the most common sensor-based threat in the context of IoT devices and applications. Sensors on IoT devices can reveal sensitive data like passwords, secret keys of a cryptographic system, credit card information, etc. This information can be used directly to violate user privacy or to build a database for future attacks. Only one sensor can be enough for information leakage (e.g., eavesdropping using microphone \cite{schlegel2011soundcomber}) or multiple sensors can be exploited to create a more complex attack (e.g., keystroke inference using gyroscope and audio sensors \cite{Narain:2014:SLK:2627393.2627417})
. In general, information leakage can be accomplished for (1) keystroke inference, (2) task inference, (3) location inference, or (4) eavesdropping as explained below. 
\subsubsection{Keystroke inference}

Keystroke inference is a generic threat on IoT devices. Most of the IoT devices provide input medium such as the touchscreen, touchpad, keyboard (external or built-in virtual or real). Whenever a user types or gives input to a device, the device tilts and turns which creates deviations in data recorded by sensors (e.g., accelerometer, gyroscope, microphone, light sensor, etc.). These deviations in sensor data can be used to infer keystrokes in an IoT device. Keystroke inference can be performed on the device itself or on a nearby device using sensors of an IoT device. 

\textit{\underline{Keystroke Inference using Light Sensors} -} Light sensors in IoT devices are usually associated with the display unit. In general, the display unit of the IoT devices is touch sensitive and provides a user interface to take inputs. For a constant state and unchangeable ambiance, the readings of the light sensor are constant. Each time a user touches and uses the touch screen to interact with the device, he/she tilts and changes the orientation of the device, which causes changes in the readings of the light sensor. Eeach input may have a dissimilar light intensity recorded by the sensor. These changes in the readings of the light sensor of a device can be utilized to infer keystrokes of that particular device. An attacker can derive the various light intensities recorded by the sensor by trying several keystroke in a device and then construnct a database. When users put their PINs or type something in the touchpad, attackers can capture the data maliciously from the device and collate these data with the database to decode keystroke information. Raphael Spreitzer developed a method named \textit{PIN Skimming} to use the data from ambient light sensor and RGBW (red, green, blue and white) sensor to extract PIN input of the smartphone \cite{Spreitzer:2014:PSE:2666620.2666622}. Markus G. Kuhn showed that input intensity changed in a cathode ray tube (CRT) display can be used to infer what is written on the screen by a photosensor of a nearby device~\cite{1004358}. In this attack, a fast photosensor of an IoT device with high-frequency components is placed in front of the CRT display to capture the signals emitted from the screen. These captured signals can be deconvoluted to reconstruct the text typed in the device. \par

\textit{\underline{Keystroke Inference using Motion Sensors} -} The main purpose of using the motion sensors (e.g., accelerometer, gyroscope, linear acceleration sensor) in IoT devices is to detect changes in motion of the devices such as shake, tilt, etc.. Accelerometer and linear acceleration sensor measure acceleration force that is applied to a device while gyroscope measures the rate of rotation in the devices. In IoT devices, the value given by the motion sensors depends on the orientation of the device and user interactions (striking force of the finger on the device display, resistance force of the hand, the location of the finger on the touchpad of the device, etc.). Thus, when a user gives inputs to a device, the motion sensors' data changes accordingly. 
Generally, IoT devices use two types of user interface to take user input -- on-screen user interface (e.g., touchpad) and external user interface (e.g., keyboard, keypad, etc.). For both user interfaces, input keys are in fixed position and for a single keystroke, the motion sensors give a specific value~\cite{cai2012practicality}. As, attackers do not need any user permission to access the motion sensors, it is easy to access the motion sensor data. 

One common keystroke inference attack can be performed by exploiting accelerometer. As mentioned above, accelerometer gives a specific reading for each user input on an IoT device, thus, attackers can build a database of pre-processed accelerometer readings with diverse input scenarios and make a matching vector of sensor data and keystrokes to extract users' input~\cite{al2013keystrokes}. The data extracted from these attacks vary from text inputs to PINs and numbers typed in the touchpad which is much more serious as attackers can acquire the PIN or credit card information~\cite{shen2015input, Aviv:2012:PAS:2420950.2420957}. 
Owusu et al. developed an app named \textit{ACCessory} which can identify the area of the touchscreen by analyzing accelerometer data of smart devices \cite{owusu2012accessory}. \textit{ACCessory} can infer PIN input on smart devices based on the detected area from accelerometer data. Accelerometer data can also be used to infer keystroke from a nearby keyboard. Marquardt et al. presented an attack scenario where accelerometer data of an IoT device can be used to guess input on a nearby keyboard \cite{Marquardt:2011:IDV:2046707.2046771}. Whenever a user types on the keyboard, a small vibration occurs and accelerometer of the IoT devices can catch this vibration and keystrokes can be identified correctly by analyzing this data~\cite{UsenixPoster}.  

Another method of keystroke inference can be achieved by analyzing gyroscope data of an IoT device. Gyroscope measures the angles of rotation in all the three axes which varies based on the specific area of the touch on the screen. Many IoT devices have a feature when users input something on the touchpad the device vibrates and gyroscope is also sensitive to this vibrational force. The orientation angle recorded in the gyroscope and the vibration caused by the input can be used to distinguish different inputs given by the users. Moreover, the data of the gyroscope can be combined with the tap sound of each key recorded via the microphone which can increase the accuracy of inferring keystrokes \cite{Narain:2014:SLK:2627393.2627417, Cai:2011:TIK:2028040.2028049}.
The combination of accelerometer and gyroscope data can also be used for keystroke inference which yields more accurate results \cite{Xu:2012:TIU:2185448.2185465, Miluzzo:2012:TYF:2307636.2307666, 7113464}.

In most wearables (smart bands, smartwatches, etc.), the motion sensors are utilized for monitoring the movement of the devices. A smartwatch, which is one of the most common wearables, maintains constant connectivity with smartphones via Bluetooth. While wearing a smartwatch, if a user moves his/her hands from an initial position, the motion sensor calculates the deviation and provides the data regarding the change of the position of the smartwatch~\cite{7382120}. Typing in the touchpad of an IoT device while wearing a smartwatch will change the data recorded by the motion sensors of the smartwatch depending on user gestures. For a specific user input interface such as \textit{QWERTY} keyboard of smartphones which has specific distance between keys, the motion sensors' data of the smartwatch can be used to infer the keystrokes~\cite{liu2015good, wang2015mole, maiti2015smart, 7368569}. 

\textit{\underline{Keystroke Inference using Audio Sensors} -} High precision microphones used in IoT devices can sense the acoustic signals emanating from keyboards (built-in or nearby) which can be used to infer the keystrokes on an IoT device. Asonov et al. proposed an experiment to record the sound of key tapping and infer the correct key from it \cite{1301311}. In this experiment, the attacker is assumed to record the acoustic signal emanating from the device while the user types on the keyboard. Then, the attacker matches this signal with a training dataset recorded stealthily while the same user was typing in the training period. 

Zhuang et al. showed that it would be possible to infer keystrokes by just analyzing the acoustic emanation without having a training data set \cite{Zhuang:2009:KAE:1609956.1609959}. In this attack scenario, a different key is assigned to a pre-defined class according to the frequency of the acoustic signal it generates while being typed. The attacker then takes a ten-minute of recording of the acoustic signal of typing on a keyboard. This recorded signal is analyzed using machine learning and speech recognition feature named \textit{Cepstrum} to match with the previously defined key classes and infer the input of a keyboard. 

In another work, Halevi et al. introduced a new technique named \textit{Time-Frequency Decoding} to improve the accuracy of keystroke inference from the acoustic signal \cite{Halevi:2012:CLK:2414456.2414509}. In this technique, machine learning and the frequency-based calculations are combined to match the recorded acoustic signal data from an IoT device with a training dataset and increase the accuracy of the attack scenario. This technique also considers typing style of users to minimize the error rate of keystroke inference. 

Berger et al. divided a PC keyboard in regions based on tap sound generated by keys and modeled a dictionary attack \cite{Berger:2006:DAU:1180405.1180436}. This attack utilizes signal processing and cross-correlation functions to process acoustic signal emanations from a nearby keyboard. Kune et al. proposed a timing attack on a number pad used in smart phone and ATMs using the audio feedback beeps generated while entering PIN \cite{FooKune:2010:TAP:1866307.1866395}. Inter-keystroke timing and distance between the numbers on the keypad are the main two features which are used to infer the input PIN in this attack. By analyzing the audio feedback recorded using microphone of a nearby IoT device, these two features are extracted and using Hidden Markov Model, the input numbers and PINs are inferred. 

Backes et al. showed that acoustic signal emanated from a dot matrix printer which was collected by a nearby microphone of an IoT device can be analyzed to predict the text printed on a paper \cite{backes2010acoustic}. In the training phase of this attack, words from a list are being printed, the acoustic signal is recorded and the data is stored. The audio signal processing and speech recognition techniques are used to extract the features of the acoustic signal to create a correlation between the number of needles used in the printer and the intensity of the audio signal. In the real attack scenario, the audio signal is captured by a nearby audio sensor and matched with previous dataset to infer the printed text. 

Zhu et al. showed a context-free attack scenario using the keyboard\textquotesingle s acoustic emanation recorded in a smartphone to infer keystrokes \cite{Zhu:2014:CAU:2660267.2660296}. In this attack scenario, the acoustic signals emanated from the keyboards are recorded by two or more smartphones. For each pair of microphones of smartphones, the recorded acoustic signal strength will depend on the distance between the typed key and the smartphones. By calculating the time-difference of the arrival of the acoustic signal, the position of the key can be inferred. 

In a similar attack, Chhetri et al. introduced a method to reconstruct the design source code sent to a 3-D printer~\cite{chhetri2016poster}. In this attack scenario, the acoustic signal emanated the 3-D printer is being recorded by a recorder placed in a close proximity of a 3-D printer and the recorded file is processed for extracting time and frequency domain features. These features are then cross-matched with a training dataset collected in a learning phase to infer the correct design. \par

\textit{\underline{Keystroke Inference using Video Sensors} -} Modern IoT devices come with powerful cameras which can both take still pictures and record high definition videos. By applying image processing techniques in captured images, keystroke inference can be done. Simon et al. developed a malware named \textit{PIN skimmer} which uses the front camera of a smartphone and microphone to infer PIN input in a smartphone \cite{Simon:2013:PSI:2516760.2516770}. PIN skimmer records the tap sound on the touchpad of a smartphone and records video using front camera of the phone. The movement recorded in the video is then analyzed to detect which part of the touchscreen is used. This information is then combined with the tap sound to infer the inputs correctly. 

Another potential malware attack to the IoT devices using the camera is \textit{Juice Filming Attack} \cite{Meng:2015:CMI:2732198.2732205}. In this attack scenario, a malicious app uses the camera to take screenshots when any user-input is given in the touchpad and save the images on storage unit (internal ROM or external memory card) of the device. Most of the IoT devices use USB for heterogeneous applications (e.g., charging, data transfer, etc.) and when the compromised device is connected to the laptop or any other device with a storage unit, the app transfers the stored pictures to the storage device from which attackers can easily extract the information. 

Shukla et al. showed a method to infer the PIN input by analyzing the hand position using the recorded video \cite{Shukla:2014:BYH:2660267.2660360}. In this method, a background application gets access to camera of smartphone and records a video when a user starts typing in a touchpad. Then, analyzing the hand position and the position of the smartphone, an attacker can extract the inputs given in a touchpad. Another version of this attack is to record the typing scenario using an external camera. In this scenario, a camera of an IoT device (e.g., smartphone, smart glass, smart surveillance system, etc.) is used to record the video of typing the PIN. In both cases, the input PIN can be inferred with high accuracy. 

Adam J. Aviv introduced another type of attack named \textit{Smudge Attack} using an external camera to infer pattern lock of an IoT device \cite{aviv2012side}. In this attack scenario, an IoT device is placed in between two cameras of other IoT devices (smartphone or smart glass) and high definition pictures are taken. Whenever the user gives the unlock pattern in the touchpad, some smudge marks are left on the screen, and captured by the cameras, which leak information about the unlock pattern to an attacker. 

Raguram et al. developed a process named \textit{iSpy} which can reconstruct the typed text by analyzing the reflection of the touchscreen in a reflective surface such as sunglass or smart glass \cite{Raguram:2011:IAR:2046707.2046769}. The experimental setup of \textit{iSpy} includes a high definition camera which can capture the video of the reflective surface while a user types in the touchpad of a phone. The reflection of the phone is being extracted from the video and consecutive frames are analyzed to extract stable pictures of the phone screen. Features (hand position, motion in the screen, etc.) are extracted from stable pictures extracted from the video and 
by using machine learning techniques, key press detection is done and typed text can be inferred successfully. \par

\textit{\underline{Keystroke Inference using Magnetic Sensors }-} Besides the aforementioned attack scenarios, electromagnetic emanations from the keyboard can be used to infer the input of a computer. As magnetic sensors of IoT devices are sensitive to electromagnetic emanations, they can be used as the attack medium. Vuagnoux et al. showed that both wired and wireless keyboards emit electromagnetic signals when a user types and this signal can be further processed to infer keystroke \cite{vuagnoux2009compromising}. In this method, electromagnetic radiation is measured by magnetic sensor of an IoT device when a key is pressed and using the  \textit{falling edge transition technique}, an attacker can infer the keystrokes. \par

\subsubsection{Task inference}
Task inference refers to a type of attack which reveals the information of an ongoing task or an application in an IoT device. Task inference reveals information about the state of the device and attackers can replicate this device state to launch an attack without alerting security policies implemented in the device. Sensors associated with IoT devices show deviation in the reading for various tasks running on the devices. This deviation in the reading can be used to infer the running process inside a device and application of the device. \par

\textit{\underline{Task Inference using Magnetic Sensors} -} Magnetic sensors in IoT devices have the role to fix the orientation of the device with respect to Earth\textquotesingle s magnetic field. Data recorded by a magnetic sensor change in the presence of an external magnetic field in the device\textquotesingle s peripheral. This deviation in data can be used to identify the tasks running on a device. Many IoT devices have a storage unit and whenever data is written or read from this storage unit, a change in the reading of magnetic sensor can be observed. Magnetic sensors of an IoT device can be used not only to infer information of the device itself, but can also be used as a medium to fetch information from a nearby device. Biedermann et al. showed that magnetic sensor of a smartphone could be used to infer on-going tasks in a storage unit like the hard drives of the computers and servers \cite{magnetic}. When an application is running on a computer, the hard drives generate a magnetic field which can be sensed by a magnetic sensor of a smartphone. Different actions cause specific readings on the magnetic sensor which can be used to track the users' action. This can be considered as a serious threat to the device and attackers can fetch valuable information in this way. \par
An electromagnetic (EM) emanation is a common phenomenon for IoT devices. Electromagnetic emanations occur whenever current passes through a device and a task is running on a device. EM emanation attacks can also be observed in  FPGA\footnote{Field-programmable gate array.}-based IoT devices \cite{quisquater2001electromagnetic, carlier2004electromagnetic, EM}. Attackers can record electromagnetic emission data generated from the FPGA-based IoT devices to deduct which kind of application is running in the system and also the states of logic blocks of the devices. Such information leakages make the system vulnerable to the user. Smart cards also emit EM waves while performing various tasks which can be captured by a radio frequency (RF) antenna and task can be inferred from the radiation \cite{6676458}. \par

\textit{\underline{Task Inference using Power Analysis} -} Power analysis is a form of sensor-based threat where an attacker studies the power consumption and power traces of the sensors for extracting information from the devices \cite{ors2003power}. O'Flynn et al. introduced an attack scenario where the power analysis attack is launched against IEEE 802.15.4 nodes \cite{eprint-2015-26786}, which is a standard low power wireless protocol used in IoT devices. Low power IoT devices use this protocol standard for various communication purposes such as connecting to a network, communicating with other devices, etc.. In this attack scenario, an attacker uses differential power analysis in the sensors. As packets transmitted from the IoT devices are encrypted, power analysis on the sensors can infer which encryption process is running in the device. Again, diverse encryption process leads to diverse power profiles which reveal associated information (e.g., key size, block size, etc.) about encryption process. Encryption process also depends on the packet size which can be observed in the power profile and attackers can infer what type of information is being transmitted based on the packet size. \par

\subsubsection{Location Inference}

Researcher developed a novel location-privacy attack based on acoustic side-channels \cite{anonymous}. The attack is based on acoustic information embedded within foreground-audio disseminated in a close environment (i.e., conference room). The researchers studied how audio, generated by secure messaging clients in voice-call
mode, can be abused to generate a location fingerprint. The
attack leverages the pattern of acoustic reflections of human voice at the user\textquotesingle s location and does not depend on any characteristic
background sounds. The attack can be used to compromise location privacy of participants of an anonymous VoIP session, or, even to carry out confirmation attacks that verify if a pair of
audio recordings originated from same location regardless of the speakers. Other researchers have also shown that several heuristics can be used to identify sensitive locations (i.e., home and work locations) of a victim whose personal device is under an adversary control \cite{petracca2016agility}.

\subsubsection{Eavesdropping}
Many IoT devices use audio sensors for making calls, recording audio messages, receiving voice commands, etc. Eavesdropping refers to a type of attack where a malicious app records a conversation stealthily by exploiting audio sensors and extract information from the conversation. An attacker can save the recorded conversation on a device or listen to the conversation in real-time. One of the recent example of eavesdropping via the microphone of a smartphone is \textit{Soundcomber} \cite{schlegel2011soundcomber}. In this example, a malicious app covertly records when a conversation is initiated from the device. As the recording is done in the background, a user does not have any idea about the recording. Several companies like banks, social security office, credit card companies, etc. have automated voice messaging system and users have to say their private information such as credit card number or social security number at the beginning of the call. Thus, \textit{Soundcomber} does not have to record all the conversation to extract data. Only the beginning part of the conversation will be enough for extracting private information of the user. Moreover, a specific conversation can also be recorded by identifying the dialed number on a smartphone. The touchpad of the smartphone creates corresponding tones when any number is dialed. This tone can be recorded and processed to identify the dialed number. After that, when a desired number is dialed, the conversation can be recorded and then processed to extract information. 

Another way to exploit microphones is to attack through voice assistant apps, e.g., Apple's Siri and Google Voice Search. Most of the IoT devices (smartphone, smart home automation, smartwatches, etc.) nowadays have built-in voice search apps. Diao et al. developed a malware named \textit{VoicEmployer} which can be installed on the device to record the voice command given in a smartphone \cite{Diao:2014:YVA:2666620.2666623}. This malware can use the recorded command for various malicious activities such as replicate malicious voice command, transfer information to paired devices, etc. \textit{Cyber Physical Voice privacy Theft Trojan horse (CPVT)} is another malware which uses the microphone of smartphones to record conversations \cite{6680832}. The recording of conversation can be controlled by external control channels like SMS, Wi-Fi, or Sensory channels \cite{6997498}. An attacker can trigger \textit{CPVT} and create command about when to start recording and when to stop recording using SMS, Wi-FI, or even sensors. Recorded conversations are stored in the device and the attacker can gain the stored files using Email, SMS, or connecting via USB. Carlini et al. showed that it is possible to exploit voice assistant apps by inserting hidden voice commands \cite{197215}. In this attack, the attacker first records voice commands of the user and extracts features from the recorded audio clips. From the extracted features, new command is generated which is not understandable by humans, but recognized by the voice assistant apps. \par

The gyroscope on IoT devices is also sensitive to an acoustic signal. Typical sampling rate of gyroscope covers some frequency of audible range which can be used to reconstruct the speech of a user. Michalevsky et al. proposed a new way of eavesdropping by analyzing vibrational noise in gyroscope caused by an acoustic signal \cite{184479}. As gyroscope does not cover the full audible range, this new process can distinguish speakers and one-syllable words by using signal processing and machine learning techniques.

\subsection{Transmitting Malicious Sensor Commands}
Sensors available in the IoT devices can be used to transmit malicious sensor patterns or triggering commands to activate malware that may have been implanted in a victim's device~\cite{6997498}. Sensors may be employed to create unexpected communication channels between device peripherals. Such channels can be used to change critical sensor parameters (e.g., devices' motion, light intensity, magnetic field, etc.) or to transmit malicious commands. \par

\textit{\underline{Transmitting via Light Sensors} -} Light sensors can be used as a potential method of transmitting signals and malicious commands \cite{joy2011side}. It is easier to transfer a bit stream via a light source by turning it on and off. Since the light sensor of an IoT device can distinguish the intensity of the light source, the light intensity change can be decoded as a bit stream in the device. By controlling the voltage of a light source, an attacker can easily transfer trigger messages and can activate malware implanted in a device. Hasan et al. showed that TV screen or laptop monitor could also be used to transfer trigger messages to a compromised IoT device by changing the light intensity of the monitor \cite{Hasan:2013:SCH:2484313.2484373}. \par

\textit{\underline{Transmitting via Magnetic Sensors} -} As mentioned earlier, magnetic sensors of an IoT device are sensitive to the magnetic fields of the device\textquotesingle s peripherals. By changing the magnetic field of the device ambiance, one can easily change the readings of the magnetic sensor which can be used as a triggering message of malware. Triggering messages encoded by an electromagnet can be sent to an IoT device and there will be some deviations in the magnetic sensor\textquotesingle s readings of the device due to this message. These deviations can be calculated and the triggering message can be extracted from this electro-magnetic signal. Moreover, the magnetic field deviations can be calculated in x, y, and z-axis and divergent values of the magnetic field deviations can be interpreted as disparate triggering messages \cite{Hasan:2013:SCH:2484313.2484373}. \par 

\textit{\underline{Transmitting via Audio Sensors} -} Audio sensors can be used to transmit malicious commands to activate a malicious application in an IoT device. Hasan et al. showed that a triggering message embedded in an audio song can be detected by the microphone and can trigger a malicious app in a smartphone \cite{Hasan:2013:SCH:2484313.2484373}. Moreover, microphones used in the modern IoT devices can detect audio signals with a frequency lower than audible range. Malware can be transferred using this audio channel as a covert channel to bypass the security measures of the device. Deshotels et al. showed that the ultrasonic sound could be used to send information to smartphones without alerting the user or any security measurement implemented on the device \cite{185183}. Subramanian et al. showed that a trojan can be transferred by encoding it in an audio signal and transferring it using a buzzer \cite{6654855}. 
\begin{table*}[t]
\centering
\caption{Summary of Existing Sensor-based Threats on IoT Devices.}
\resizebox{\textwidth}{!}{
\begin{tabular}{| g | c | c | c | c | c | c | c | c |}\hline
\rowcolor[HTML]{C0C0C0}
\backslashbox{Threats}{Sensors}   & \begin{tabular}[c]{@{}c@{}}Light\\Sensor\end{tabular} & \begin{tabular}[c]{@{}c@{}}Motion\\Sensor\end{tabular} & \begin{tabular}[c]{@{}c@{}}Magnetic\\Sensor\end{tabular} & \begin{tabular}[c]{@{}c@{}}Acoustic\\Sensor\end{tabular} & GPS & Camera & \begin{tabular}[c]{@{}c@{}}Power\\Analysis\end{tabular}\\ \hline \hline
 Information Leakage & \cite{Spreitzer:2014:PSE:2666620.2666622}, \cite{1004358} & \makecell{\cite{cai2012practicality}, \cite{al2013keystrokes}, \cite{shen2015input}, \cite{Aviv:2012:PAS:2420950.2420957},\\ \cite{7382120}, \cite{owusu2012accessory}, \cite{Marquardt:2011:IDV:2046707.2046771}, \cite{Narain:2014:SLK:2627393.2627417}, \\ \cite{Cai:2011:TIK:2028040.2028049}, \cite{184479}, \cite{Xu:2012:TIU:2185448.2185465},\\ \cite{Miluzzo:2012:TYF:2307636.2307666}, \cite{7113464}, \cite{liu2015good},\\ \cite{wang2015mole}, \cite{maiti2015smart}, \cite{7368569}} & \makecell{\cite{magnetic}, \cite{vuagnoux2009compromising}, \cite{quisquater2001electromagnetic}, \\ \cite{carlier2004electromagnetic}, \cite{EM}, \cite{6676458}} & \makecell{\cite{1301311}, \cite{Zhuang:2009:KAE:1609956.1609959}, \cite{Halevi:2012:CLK:2414456.2414509}, \cite{Berger:2006:DAU:1180405.1180436}, \\ \cite{FooKune:2010:TAP:1866307.1866395}, \cite{backes2010acoustic}, \cite{Zhu:2014:CAU:2660267.2660296},  \cite{schlegel2011soundcomber}, \\ \cite{Diao:2014:YVA:2666620.2666623}, \cite{6680832}, \cite{6997498}, \cite{chhetri2016poster},\\  \cite{petracca2017aware}, \cite{Petracca:2015:APA:2818000.2818005}} &  \cite{petracca2016agility} & \makecell{\cite{Simon:2013:PSI:2516760.2516770}, \cite{Meng:2015:CMI:2732198.2732205}, \cite{Shukla:2014:BYH:2660267.2660360},\\ \cite{aviv2012side}, \cite{Raguram:2011:IAR:2046707.2046769}, \cite{petracca2017aware}} & \cite{ors2003power}, \cite{eprint-2015-26786} \\ \hline
     \makecell{Transferring Malware\\or Malicious Code} & \cite{Hasan:2013:SCH:2484313.2484373}, \cite{joy2011side} & --- & \cite{Hasan:2013:SCH:2484313.2484373} & \cite{Hasan:2013:SCH:2484313.2484373}, \cite{6654855}, \cite{185183}  & --- & --- & --- \\ \hline
     False Data Injection & --- & --- & --- & ---  & \cite{tippenhauer2011requirements}, \cite{coffed2014threat}, \cite{Giannetsos:2013:SST:2463183.2463186} & --- & \cite{7868342} \\ \hline
     Denial-of-Service & \cite{ICS-CERT} & \cite{son2015rocking} & --- & ---  & --- & --- & --- \\ \hline
\end{tabular}}
\end{table*}
\subsection{False Sensor Data Injection}
The applications of IoT devices largely depend on data collected by sensors available on the devices. By altering the sensor data, one can control the applications of IoT devices. False sensor data injection refers to an attack where the sensor data used in the IoT applications is forged or forcefully changed to perform malicious activities. The false sensor data can be injected in the devices by accessing the device physically or by using various communication medium (Bluetooth, Wi-Fi, cellular network, etc.) covertly. Moreover, the sensors of IoT devices can also be used to alter data typed or stored on the devices. 

Tippenhauer et al. showed a \textit{spoof attack} scenario in GPS-enabled devices to change the real location of the device \cite{tippenhauer2011requirements}. In this attack scenario, a vehicle with a GPS enabled device is used. Attacker transmits a forged GPS signal to the device to alter the location of the vehicle. In this way, the real location of the vehicle is disguised and the attacker can perform any physical attack to the disguised vehicle. The GPS data used in the smartwatches can expose the location of a user and this GPS data can then be forged and a new location can be given as a false input in the GPS \cite{coffed2014threat}. 

The power analysis attack on IoT devices can also be used for injecting false data. The power analysis on IoT devices running an encryption algorithm can reveal information about encryption process including the block size, key size, even the actual encryption key \cite{7868342}. This information can be used to encrypt a false data and replace the original data on the device. Thus, attackers can inject false encrypted data in the communication channel to change the action of a device for specific commands. 

Giannetsos et al. introduced a malicious app named \textit{Spy-sense}, which monitors the behavior of the sensors in a device and can manipulate data by deleting or modifying it \cite{Giannetsos:2013:SST:2463183.2463186}. \textit{Spy-sense} exploits the active memory region of a device and alters the data structure and reports back important data to a server covertly.

\subsection{Denial-of-Service}
Denial-of-Service (DoS), by definition, is a type of attack where the normal operation of a device or application is denied maliciously. DoS attacks can be active attacks where an application or task is refused forcefully or passive attacks where attacking one application can stop another on-going task on the device. Recently, ICS-CERT published an active alert for a list of accelerometers used in IoT devices which can be exploited using vibrational force \cite{ICS-CERT}. Every accelerometer has a working frequency and if an external vibrational force can match this frequency, it is possible to turn off the devices forcefully. Son et al. showed that it is possible to obstruct the flight control of a drone by exploiting gyroscope using a sound signal \cite{son2015rocking}. The MEMS Gyroscopes deployed in drones have a sensing mass inside of the sensor which is constantly vibrating. The gyroscope measures the rotational motion of the device with respect to the sensing mass. When the resonant frequency of the gyroscope is matched by an audio signal, an attacker can obstruct the normal performance of the gyroscope and change the course of the drone, or even turn it off. 

\section{Existing Security Mechanisms to Prevent Sensor-based Threats}

Researchers have identified a diverse set of sensor-based threats for IoT devices. Table 2
lists a summary of existing sensor-based threats on IoT devices. Although there are several threats, no comprehensive security mechanism able to prevent such threats has been developed yet. Indeed, the use of a wide range of sensors in IoT devices and applications has made it hard to secure all the sensors by one effective framework. Furthermore, the lack of knowledge of the existing sensor-based threats and differences in sensor characteristics make it hard to establish a complete and comprehensive security measure to secure all the sensors of IoT devices against the sensor-based threats \cite{203854}. \par

In this section, we discuss two main approaches proposed by researchers in an attempt to design security mechanisms for sensor-based threats on IoT devices.

\textbf{\textit{Enhancing Existing Sensor Management Systems}}. One approach toward securing the sensors in IoT devices is to enhance existing sensor management systems of IoT OSes. For instance, Xu et al. proposed an extension of the Android sensor management system named \textit{Semadroid}, which provides users with a monitoring and logging feature to make the usage of sensors by apps explicit. Also, with \textit{Semadroid}, users can specify policies to control whether and with what level of precision third party apps can access to sensed data. Moreover, \textit{Semadroid} creates mock data to verify how applications, from unknown vendors, use sensed data and, thus, prevents malicious behaviors. \par
Furthermore, system designers have long struggled with the challenge of determining how to let the user control when applications may perform operations using privacy-sensitive sensors securely and effectively. Current commercial systems request that users authorize such operations once (i.e., on install or first use), but malicious apps may abuse such authorizations to collect data stealthily using such sensors. Proposed research methods enable systems to infer the operations associated with user input events \cite{onarlioglu2016overhaul, roesner2012user, ringer2016audacious}, but malicious applications may still trick users into allowing unexpected, stealthy operations. To prevent users from being tricked, Petracca et al. proposed to bind applications' operation requests to the associated user input events and how such events are obtained explicitly, enabling users to authorize operations on privacy-sensitive sensors unambiguously \cite{petracca2016aware, petracca2017aware}. To demonstrate this solution, they implemented the \textit{AWare} authorization framework for Android, extending the Android Middleware to control access to privacy-sensitive sensors. They evaluated the effectiveness of \textit{AWare} in: (1) a laboratory-based user study, finding that at most 7\% of the users were tricked by examples of four types of attacks when using \textit{AWare}, instead of 85\% on average for prior approaches; (2) a field study, showing that the user authorization effort increases by only 2.28 decisions on average per application; (3) a compatibility study with 1,000 of the most-downloaded Android apps, demonstrating that such applications can operate effectively under \textit{AWare}. Moreover, an alternative mechanism is proposed in \textit{6thSense}, where researchers proposed a context-aware framework to detect the sensor-based threats in IoT devices~\cite{203854}. This framework is built upon the observation that for any user activity on an IoT device, a specific set of sensors becomes active. \textit{6thSense} builds a comprehensive context-aware model for each user activity based on this observation. Differently from other works, \textit{6thSense} utilizes all the sensor data in real-time and determines whether the present context of the sensors is malicious or not using various machine learning-based approaches. Researchers tested the proposed framework with 50 real-life user data and confirmed that \textit{6thSense} can detect various sensor-based threats with approximately 97\% accuracy and F-score.

\textbf{\textit{Protecting Sensed Data}}. Another approach toward securing IoT devices against the sensor-based threats is to protect the sensed data in transfer and at rest. Indeed, some malicious applications record sensor data and transmit it later when the device is locked or when security protection mechanisms are turned off. For instance, sensed location data may be subject to inference attacks by cybercriminals that aim to obtain sensitive locations such as the victim's home  and work locations to launch a variety of attacks. 

Location-Privacy Preserving Mechanisms (LPPMs) exist to reduce the probability of success of  inference attacks on location data. However, such mechanisms have been shown to be less effective when the adversary is informed of the protection mechanism adopted, also known as \textit{ white-box attacks}. Petracca et al. proposed  a  novel approach that makes use of targeted maneuvers to augment real sensors' data with synthetic data and obtain a uniform distribution of data points, which creates a robust defense against \textit{white-box attacks} \cite{petracca2016agility}. Such maneuvers are systematically activated in response to specific system events to rapidly and continuously control the rate of change in system configurations and increase  diversity in the space of readings, which would decrease the probability of  success of inference attacks by an adversary. Experimental results performed on a real data set showed that the adoption of such maneuvers reduces the probability of success of white-box attacks to 3\% on average compared to 57\% when using the state-of-the-art LPPMs.

Furthermore, power analysis attacks and electromagnetic emanation attacks exploit information from the power consumption and electromagnetic emissions of active sensors from the device. One proposed countermeasure to immune electromagnetic emanation attacks is to use a single inverter ring oscillator (SIRO) \cite{zafar2008novel}. In this proposed system, a multi-clock system with cipher embodiment is used with SIRO-based synchronization. The absence of external oscillator and unsynchronized nature of SIRO makes the system more immune to electromagnetic emission. Again, SIRO-based system provides frequency hopping scheme in cipher which increases immunity to timing and power analysis attacks. Standaert et al. proposed an approach to minimize the effect of power analysis attack which is based on the correlation between the power consumption measurements and a simple prediction developed on the number of bit transitions within the devices \cite{power}. The use of random pre-charges in the devices can minimize the probability of power analysis attack on the FPGA-based IoT devices.\par

More general solutions to address the protection of the sensed data have also been proposed. For example, Roman et al. proposed the use of public key encryption to secure sensor data from devices \cite{roman2011key}. They proposed the encryption of sensor data collected and stored it in the device before sharing it with third party apps or other devices. Devices connected to each other can share their public key through a key management system and use their assigned private key to decrypt the sensor data. Third party apps installed in the device can also use public key encryption scheme to use sensor data for various applications.  \par

Trust management frameworks can also be leveraged for secure information flow among sensors, secure communication of sensor data with other devices, and to certify authorized access of sensors by trusted software and apps in the system. Trust management frameworks can over-access requests on sensors and take decisions based on whether the requests are legitimate or not. For instance, a framework named \textit{AuDroid} was proposed to secure communications via audio channels when applications make use of the device's microphones and speakers \cite{Petracca:2015:APA:2818000.2818005}. \textit{AuDroid} leverages the SELinux kernel module to build a reference monitor which enforces access control policies over dynamically created audio channels. It controls information flows over audio channels and notifies users whenever an audio channel is created between processes at runtime. \par

\textbf{Shortcomings of Proposed Security Mechanisms ---} Although the aforementioned solutions address sensor-based threats, there are still limitations  that need to be overcome. 

(1) Most of the proposed security mechanisms for IoT devices are anomaly detection frameworks at the application level which are not suitable for detecting sensor-based threats at the system level \cite{Wang:2015:NHM:2802130.2802132, Sun:2014:DIA:2664243.2664245, Wu:2014:DDA:2663761.2664223, 7294263}. Sikder et al. analyzed the performance of several sensor-based threats with respect to real-life malicious software scanners available in \textit{VirusTotal} website and observed that no scanner can recognize sensor-based threats~\cite{203854}.(2) With the growing popularity of the IoT concept, more and more devices are being interconnected with each other and the security of these devices becomes difficult to manage. Many IoT devices are severely resource-limited, small devices and it is hard to implement a complex security mechanism considering the limited resources of the devices \cite{sicari2015security}. (3) Proposed security mechanisms only target a subset of sensitive sensors available in IoT devices nowadays. For instance, commercial sensor management systems use an explicit permission-based security model for only some of the sensors (e.g., camera, GPS, and microphone). Similarly, \textit{AuDroid} provides a policy-enforced framework to secure the audio sensors of IoT devices explicitly \cite{Petracca:2015:APA:2818000.2818005}; however, such framework was not designed to protected other sensitive sensors. Other proposed solutions only provide protection against power analysis and electromagnetic emanation-based attacks, respectively \cite{zafar2008novel, power}. A step forward was made with \textit{AWare} and \textit{6thSense} that covered a wider set of privacy-sensitive sensors available in current IoT devices to build a context-aware model and determine whether a sensor usage scenario is malicious. (4) In solutions where users' decisions are utilized to build the sensor use policy for third party apps, such as in \textit{Semadroid} and \textit{AWare}, if a user allows an application to use a sensor without any restriction, then the application is blindly treated as secure by the system. (5) Encrypting sensor data using public key encryption schemes provides protection to sensor data, but it also consumes high power to run in smaller IoT devices \cite{roman2011key}. This power-performance trade-off is impractical for resource-limited IoT devices. 

In conclusion, a complete and comprehensive solution for autonomous policy enforcement, comprehensive coverage of all the sensors, and an efficient power-performance trade-off is yet to be designed.

\section{Open Issues and Further Research}
The concept of IoT is no longer in the developing stage and new research ideas related to IoT are emerging these days. 
In this section, we discuss open issues and future research directions in the context of sensor-based threats. 

\par \textbf{Study of Expected Functionality to Identify Threats -} Researchers should study the functionality expected from IoT systems to identify threats. As the IoT concept is relatively new, less knowledge about the internal architectures (i.e., software and hardware) of IoT devices 
is available, which is an obstacle to secure sensors in IoT devices. Additionally, researchers and users know less about sensor-based threats which are lucrative for attackers to target IoT devices~\cite{sadeghi2015security}.  
Users carelessly install any third party apps with illegitimate sensor permissions which can compromise IoT devices~\cite{felt2012android, felt2012ask}. Therefore, to secure sensors in IoT devices, it is important to understand how users are using the devices and what their views of sensor-based threats are. Researchers may perform additional usability studies to better understand how users can contribute to improve sensor access control via their inputs in IoT devices. 


\par
\textbf{Adoption of Standard Security Mechanisms -} 
Currently, there exist several operating systems for IoT devices that manage their on-board sensors in dissimilar ways. These dissimilarities make it hard to converge for a general security scheme to protect sensors of the IoT devices~\cite{mobilewrap}. One of the future research efforts should be the standardization of development platforms for IoT devices which will make it easier for researchers to come up with universal security measures to defend against sensor-based threats. Therefore, researchers should investigate the possibility of a common security mechanism for authentication of sensor data as well as authorization of legitimate sensor access. 


\par \textbf{Fine-grained Control of Sensors -} Existing sensor management systems of IoT devices offer permission-based sensor management which completely depends on user consent. Apps generally ask for permissions to access specific sensors on installation time and once the permissions are granted, users have less control over the sensors' usage by the apps. Again, the user permission is enforced only to secure limited number of the on-board sensors (e.g., microphone, camera, GPS). Granting permission to these sensors automatically grant permission for other sensors such as accelerometer, gyroscope, light sensor, etc. In addition, in recent years, researchers have also showed that both permission-enforced (microphone, camera, GPS) and no permission-enforced (accelerometer, gyroscope, light sensor, etc.) sensors are vulnerable to sensor-based threats.  Therefore, a fine-grained sensor management system is needed to verify compliance between sensor access and user intent.

\par \textbf{Control Sharing of Data among Sensors -} Communication on IoT devices become more sensor-to-sensor (i.e., machine-to-machine) compared to human-to-sensor or sensor-to-human (human-to-machine or machine-to-human) and the introduction of huge number of sensors in IoT devices is speeding up this shift. As IoT devices deal with sensitive personal data,  sensor-to-sensor communication channels should be secured, which helps in end-to-end security for the devices. Secure end-to-end communication from sensors to the devices and among devices are vital to avoid information leakage \cite{hossain2015towards, Weber201023}.

\par \textbf{Protect Sensor Data when at Rest -} IoT applications deal with multiple sensor data at a time and tampered data in the IoT devices can impact the normal behavior of applications. To ensure authenticity of sensor data, various  
encryption mechanisms may be applied from the sensors to the program requesting it. Different security features of the hardware such as ARM TrustZone may be adopted to achieve secure data flow inside the devices~\cite{namiluko2013towards}. Researchers may also invest their effort in studying the adoption of the blockchain technology as a way of designing highly distributed systems able to provide attestation and verification among multiparty and heterogeneous components part of a larger IoT system. 

\par \textbf{Prevent Leakage of Secret Data -} IoT devices can autonomously sense their surrounding environment which can be used to prevent information leakage from the devices. Sensors in IoT devices can anticipate an ongoing task and detect pattern of information accessed by the task. These sensor patterns varies for different activities and by observing these sensor behaviors, it is possible to prevent information leakage in IoT devices~\cite{203854}.

\par \textbf{Protect Integrity of Sensor Operations -} The research community has not invested enough effort in studying the design and development of tools for automated detection and analysis of sensors-based threats. For instance, no tool is available to automatically identify and analyze adversary-controlled sensors that would compromise the integrity of sensor operations, as well as, the integrity of the data generated or modified by such operations. Also, no tool is available to automatically identify dangerous configurations in enforced access control policies, which may lead to risky operations by trusted programs that may compromise the integrity of the entire IoT system. 

\par \textbf{Adoption of Intrusion Mechanisms to Detect Attacks -} In recent years, multiple efficient techniques (e.g., machine learning (ML) and  
neural network (NN)) were applied to detect threats in various application domains. These detection techniques should be explored in detail to design novel intrusion detection mechanism, for IoT devices and applications, able to identify when unsafe operations are authorized. Therefore, researchers should investigate NN and ML classification algorithms as viable solutions to identify and differentiate legitimate from illegal sensing activities. 


\par \textbf{Summary -} In summary, there are several interesting research problems that may be tackled by the research community toward improving the security of sensors in IoT devices and applications. While following the above directions toward better protection mechanisms against sensor-based threats, researchers have to identify the key characteristics that differentiate IoT security from the commodity system security. Such unique characteristics may guide toward the design of innovative mechanisms never thought before.

\section{Conclusion}

The growing popularity of IoT is increasing attention towards security issues in IoT devices and applications. In this paper, we surveyed a lesser known yet serious family of threats: \textit{sensor-based threats to IoT devices}. We presented a comprehensive overview of sensors in IoT devices and existing sensor management systems adopted in commodity IoT OSes. We provided a detailed analysis of recent sensor-based threats and discussed how these threats can be used to exploits various sensors in IoT devices. We also summarized several security approaches proposed by researchers in the attempt to address critical shortcomings for the security of current IoT systems, and discussed some of the challenges for future research work in this area. In conclusion, we believe this survey will have a positive impact in the research community by documenting recent sensor-based threats to IoT devices and motivating researchers to develop comprehensive security schemes to secure IoT devices against sensor-based threats.


%

\section*{Acknowledgment}

The authors would like to thank US National Science Foundation to support this work under the award NSF-CAREER-CNS-1453647. This research was sponsored by the Army Research Laboratory and was accomplished under Cooperative Agreement Number W911NF-13-2-0045 (ARL Cyber Security CRA). The views and conclusions contained in this document are those of the authors and should not be interpreted as representing the official policies, either expressed or implied, of the Army Research Laboratory or the U.S. Government. The U.S. Government is authorized to reproduce and distribute reprints for Government purposes not with standing any copyright notation here on. 

\vspace{-0.5cm}




%
\bibliographystyle{IEEEtran}
\bibliography{references}
\vspace{-1cm}

%
\begin{IEEEbiography}[{\includegraphics[width=1in,height=2in,clip,keepaspectratio]{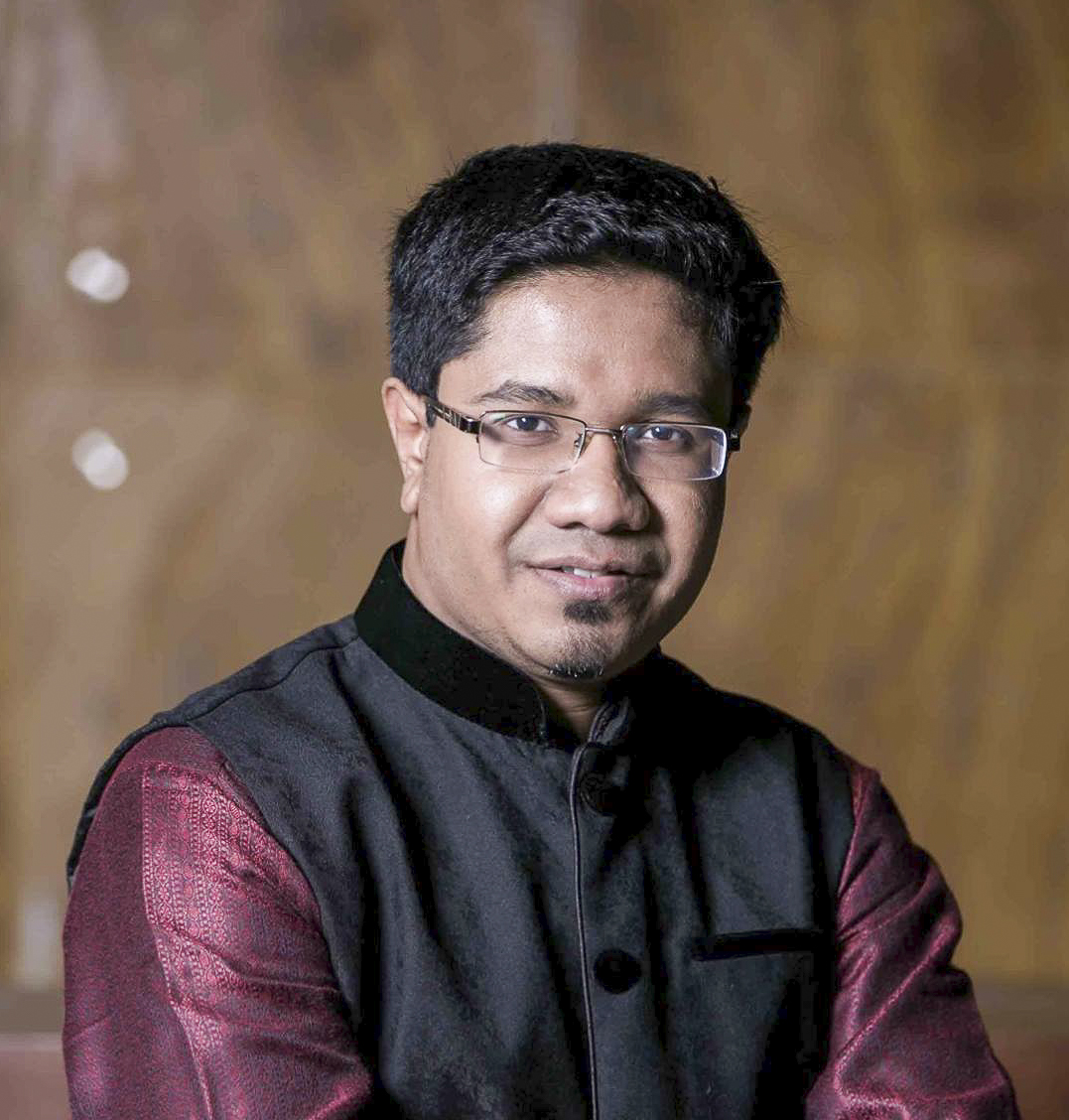}}]{Amit Kumar Sikder}

 is currently a PhD student and Research Assistant in the Department of Electrical and Computer Engineering at Florida International University, as a member of the Cyber-Physical Systems Security Lab (CSL). He previously completed his Bachelors in Electrical and Electronic Engineering from Bangladesh University of Engineering and Technology (BUET). His research interests are focused on the security of Cyber-Physical Systems (CPS) and Internet of Things (IoT). He also has worked in areas related to security of smart devices, security of smart home, smart city, wireless communication. More information can be obtained from: http://web.eng.fiu.edu/asikd003/. 
\end{IEEEbiography}
\vspace{-1cm}

\begin{IEEEbiography}[{\includegraphics[width=1in,height=2in,clip,keepaspectratio]{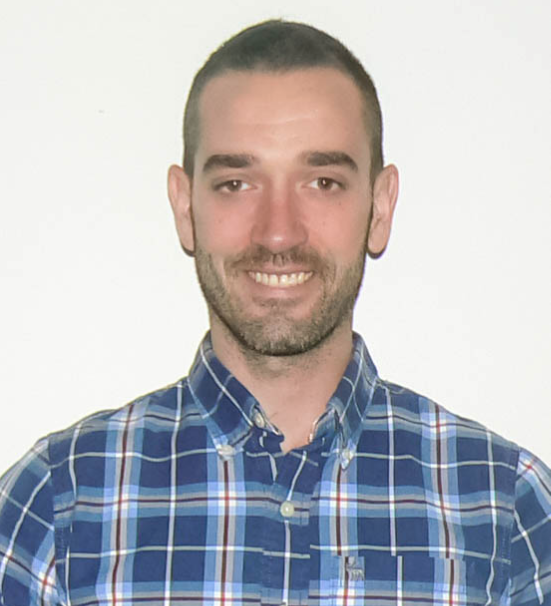}}]{Giuseppe Petracca}
 is currently a PhD student and Research Assistant in the Department of Computer Science and Engineering at The Pennsylvania State University. He also collaborates for the Cyber Security Collaborative Research Alliance (CRA), sponsored by the Army Research Laboratory (ARL). Giuseppe has a B.S. and a M.S. in Computer Science and Engineering from Sapienza University of Rome, Italy. Giuseppe's research interest focuses on mobile systems and cloud computing security. His industry experience includes a summer internship in 2013 as Graduate Researcher at Intel, a summer internship in 2014 as Graduate Technical Engineer at Intel Labs, a summer internship in 2016 as Software Engineer and Security Researcher at Samsung Research America, and a summer internship in 2017 as Software Engineer and Security Researcher at Google. More information can be obtained from: http://sites.psu.edu/petracca/. 
\end{IEEEbiography}

\begin{IEEEbiography}[{\includegraphics[width=1in,height=2in,clip,keepaspectratio]{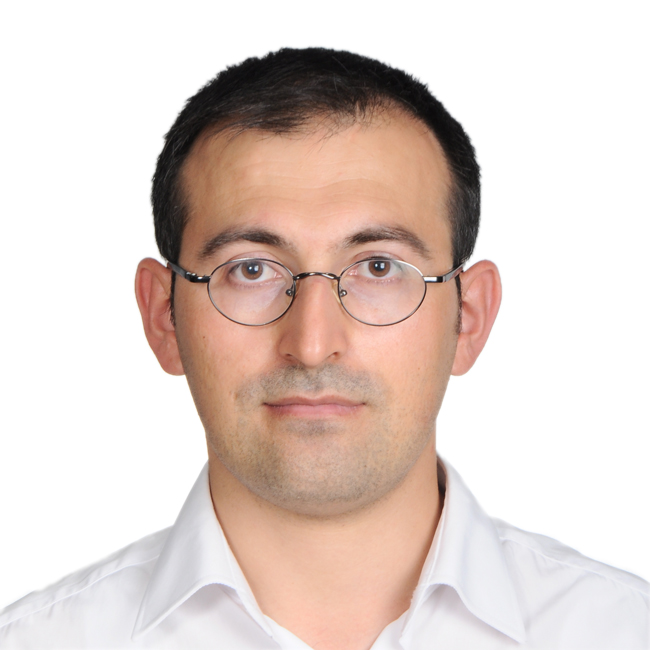}}]{Hidayet Aksu}
received his Ph.D., M.S. and B.S. degrees from Bilkent University, all in Department of Computer Engineering, in 2014, 2008 and 2005, respectively. He is currently a Postdoctoral Associate in the Department of Electrical \& Computer Engineering at Florida International University (FIU). Before that, he worked as an Adjunct Faculty in the Computer Engineering Department of Bilkent University. He conducted research as visiting scholar at IBM T.J. Watson Research Center, USA in 2012-2013. He also worked for Scientific and Technological Research Council of Turkey (TUBITAK). His research interests include security for cyber-physical systems, internet of things, security for critical infrastructure networks, IoT security, security analytics, social networks, big data analytics, distributed computing, wireless networks, wireless ad hoc and sensor networks, localization, and p2p networks.
\end{IEEEbiography}
\vspace{-3cm}

\begin{IEEEbiography}[{\includegraphics[width=1in,height=2in,clip,keepaspectratio]{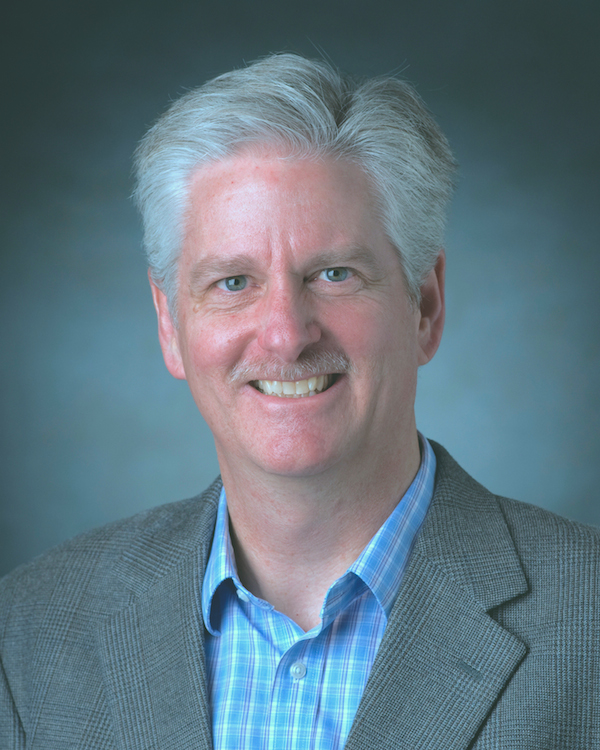}}]{Dr. Trent Jaeger}
is a Professor in the Computer Science and Engineering Department at The Pennsylvania State University and the Co-Director of
PSU's Systems and Internet Infrastructure Security (SIIS) Lab.  Trent's research interests include systems security and the application of programming language techniques to improve security.  He has published over 100 refereed papers on these topics and the book "Operating Systems Security," which examines the principles behind secure operating systems designs.  Trent has made a variety of contributions to open source systems security, particularly to the Linux Security Modules framework, SELinux, integrity measurement in Linux, and the Xen security architecture.  He was previously the Chair of the ACM Special Interest Group on Security, Audit, and Control (SIGSAC). Trent has an M.S. and a Ph.D. from the University of Michigan, Ann Arbor in Computer Science and Engineering in 1993 and 1997, respectively, and spent nine years at IBM Research prior to joining Penn State. More information can be obtained from: http://www.cse.psu.edu/\textasciitilde trj1/. 
\end{IEEEbiography}
\vspace{-3cm}

\begin{IEEEbiography}[{\includegraphics[width=1in,height=2in,clip,keepaspectratio]{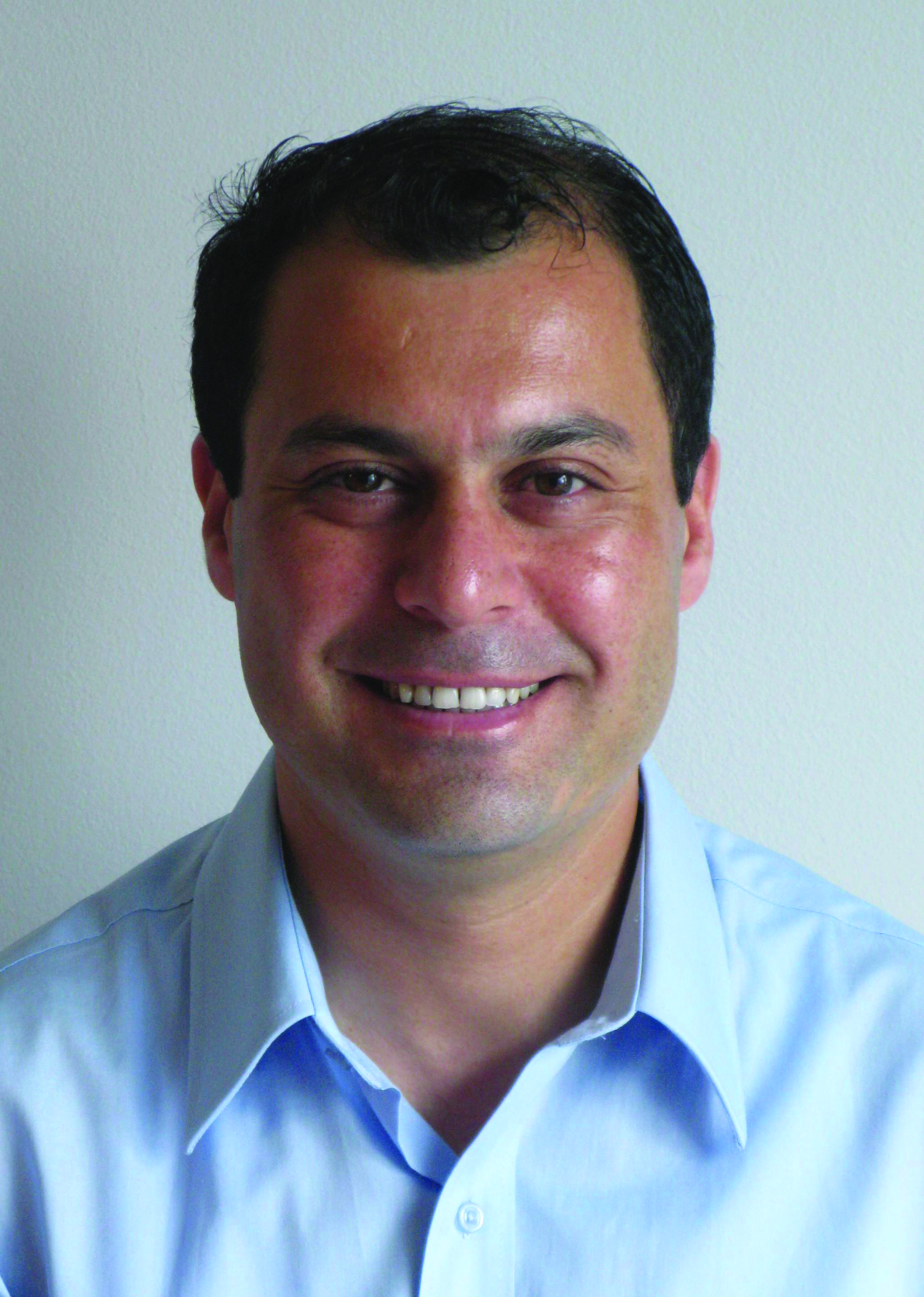}}]{Dr. A. Selcuk Uluagac} is currently an Assistant Professor in the Department of Electrical and Computer Engineering (ECE) at Florida International University (FIU). Before joining FIU, he was a Senior Research Engineer in the School of Electrical and Computer Engineering (ECE) at Georgia Institute of Technology. Prior to Georgia Tech, he was a Senior Research Engineer at Symantec. He earned his Ph.D. with a concentration in information security and networking from the School of ECE, Georgia Tech in 2010. He also received an M.Sc. in Information Security from the School of Computer Science, Georgia Tech and an M.Sc. in ECE from Carnegie Mellon University in 2009 and 2002, respectively. 
The focus of his research is on cyber security topics with an emphasis on its practical and applied aspects. He is interested in and currently working on problems pertinent to the security of Cyber-Physical Systems and Internet of Things. In 2015, he received a Faculty Early Career Development (CAREER) Award from the US National Science Foundation (NSF).
In 2015, he was awarded the US Air Force Office of Sponsored Research (AFOSR)'s 2015 Summer Faculty Fellowship. In 2016, he received the Summer Faculty Fellowship from the University of Padova, Italy.
He is also an active member of IEEE (senior grade), ACM, and ASEE and a regular contributor to national panels and leading journals and conferences in the field. Currently, he is the area editor of Elsevier Journal of Network and Computer Applications and serves on the editorial board of the IEEE Communication Surveys and Tutorials. More information can be obtained from: http://web.eng.fiu.edu/selcuk.
\end{IEEEbiography}






\end{document}